\documentclass[pra,aps,preprintnumbers,superscriptaddress]{revtex4}
\usepackage{epsfig,amsmath,amssymb,amsfonts}
\def\CC{{\rm\kern.24em \vrule width.04em height1.46ex depth-.07ex
\kern-.30em C}}
\def\P{{\rm I\kern-.25em P}}
\def\RR{{\rm
         \vrule width.04em height1.58ex depth-.0ex
         \kern-.04em R}}
\def\RR{{\rm\kern.24em \vrule width.04em height1.46ex depth-.07ex
\kern-.30em R}}
\def\P{{\rm I\kern-.25em P}}
\def\RR{{\rm
         \vrule width.04em height1.58ex depth-.0ex
         \kern-.04em R}}

\newcommand{\ket}[1]{\left | \, #1 \right\rangle}

\newcommand{\half}{\mbox{$\textstyle \frac{1}{2}$}}

\newcommand{\be}{\begin{equation}}
\newcommand{\ee}{\end{equation}}
\newcommand{\bq}{\begin{eqnarray}}
\newcommand{\eq}{\end{eqnarray}}

\begin{document}

\title{Conditional Aharonov-Bohm Phases with Double Quantum Dots}

\author{Roberta Rodriquez\footnote[1]{r.rodriquez@damtp.cam.ac.uk} }
\address{Department of Physics, Cavendish Laboratory, \\
University of Cambridge, Cambridge, CB3 0HE, UK,}
\address{Department of Applied Mathematics and Theoretical
Physics,\\ University of Cambridge, Cambridge CB3 0WA, UK.}
\author{Jiannis K. Pachos\footnote[3]{j.pachos@damtp.cam.ac.uk}
}

\address{Department of Applied Mathematics and Theoretical
Physics,\\ University of Cambridge, Cambridge CB3 0WA, UK.}

\begin{abstract}
A quantum dot proposal for the implementation of topological
quantum computation is presented. The coupling of the electron
charge to an external magnetic field via the Aharonov-Bohm effect,
combined with the control dynamics of a double dot, results in a
two-qubit control phase gate. The physical mechanisms of the
system are analysed in detail and the conditions for performing
quantum computation resilient to control errors are outlined and found
to be realisable with present technology.
\end{abstract}

\maketitle



\section{Introduction}

Solid state is becoming an increasingly promising arena for future
implementations of quantum computation
\cite{Divincenzo1,Divincenzo2,Briggs}, as a consequence of the
vast progress of classical computation technology in the atomic
scale. Quantum manipulation of solid state devices paves the way
for many proposals supporting, in principle, the possibility of
scalable quantum  computation. However, the control of individual
quantum systems, such as electrons in quantum dots, demands a much
higher degree of control accuracy than currently available. A
number of recent proposals address the issue of controllability by
employing geometrical and topological effects
\cite{Kitaev,Jiannis}. The key advantage of these methods is that
the resulting geometrical and topological gates do not depend on
the overall time of the evolution, nor on small deformations in
the control parameters. Possible manifestations of geometrical
phases are the Berry phases obtained, for example, through a
cyclic adiabatic evolution of a system \cite{Berry}, or through the
Aharonov-Bohm effect \cite{AB}. Within solid state physics, there
have even been proposals to implement Berry phases with Josephson
junctions \cite{Vlatko} as well as Aharonov-Bohm phases encoded on
the different spin states of electrons manipulated in quantum dot
structures \cite{Akera,Loss}.

In this chapter we present a simple solid state scheme for
geometrical quantum computation where the Aharonov-Bohm phase is
encoded on the electronic charge \cite{Pachos}. In particular, we
consider quantum dots that can either be empty or accommodate
one electron. In an array of quantum dots we assume that one is
able to lower the potential between any two dots and facilitate
the quantum tunneling between them. Intrinsically, the Hamiltonian
of this system is governed mainly by three terms, namely the
potential wall of height $V_0$ separating two neighbouring dots
and the Coulomb interaction $V_C=e^2/r$ between two electrons
occupying the same dot or adjacent dots in close proximity.

\begin{figure}[!htp]
\begin{center}
\includegraphics[width=12cm]{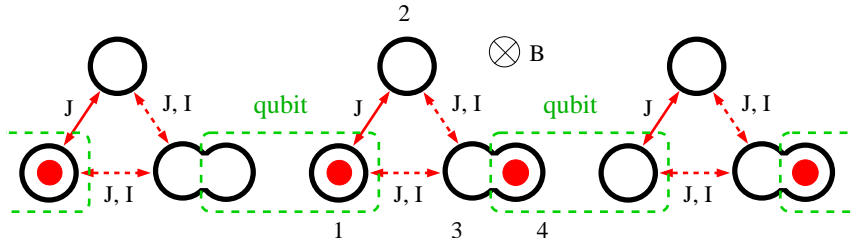}
\end{center}
\caption[chain]{Array of quantum dots illustrating the dots within
the computational space (1 and 4) and the auxiliary
dots (2 and 3). The auxiliary dots get populated only during the operation
of the two-qubit gates. Through the triangular configuration (1, 2 and 3)
a circular path is obtained. The double dot (3 and 4) provides the
conditional dynamics.}\label{fig:array}
\end{figure}

Imagine that we have an array of paired dots with one electron in
each pair, as shown in Fig. \ref{fig:array}. The position of an
electron in the pair encodes the states $|0\rangle$ and
$|1\rangle$. If the electron is in the left dot the qubit $i$ is
in the logical state $\ket{l}_i=\ket{0}$, while if the electron is in
the right dot the qubit is in state $\ket{r}_i=\ket{1}$.
Preparation of the initial qubit state as well as the final
readout are therefore technically easy tasks. Moreover,
single-qubit gates are simple to achieve in this setting \cite{fuji}. By
lowering the potential barrier $V_0$ between the pair of the dots
that compose a qubit, quantum tunnelling between the two dots will
create superpositions of the logical states \cite{Bonadeo} giving,
for example, the state $|\psi\rangle=c_0 |0\rangle + c_1
|1\rangle$. Any possible one qubit rotation can thus
be performed. For a two-qubit phase-gate we need to perform a
controlled operation where the state of one of the qubits is
changed conditionally on the state of the second qubit. We shall
present a way of performing such a gate in a topological fashion
by employing, in addition, a double dot structure. The combination
of single-qubit gates and controlled phase-gates can then allow us
to carry out any arbitrary quantum computation
\cite{Barenco_et_al}. Finally, comparison of our model with state
of the art experiments is presented together with an outline of
its potential advantages.

\section{Two qubit phase-gates}
\label{gate}

We would like to implement a control phase gate by adopting an
intrinsically geometrical evolution. Consider a homogeneous
magnetic field $B$ in the neighbourhood of the dots that comprise
the logical array of qubits. An electron which spans a closed
trajectory (loop) inside the magnetic field will acquire a phase
factor proportional to the flux of the magnetic field encircled by
the loop \cite{AB}. If this trajectory is spanned in a conditional
way, depending on the presence of an electron in a neighbouring
dot, the resulting effect can be used as the basis for a two-qubit
phase-gate.

\begin{figure}[!htp]
\begin{center}
\includegraphics[width=4cm]{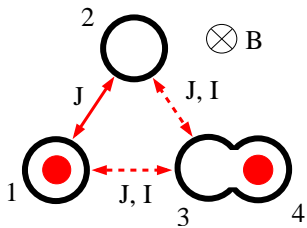}
\end{center}
\caption[chain]{Quantum dot implementation of the traversing loop
by tunnelling transitions in the presence of a magnetic field. The
possible occupancy of dot 4 changes the tunneling coupling towards dot
3 from $J$ to $I \ll J$ eventually forbidding the circulation of the
electron in dot 1.}\label{fig:loop}
\end{figure}

We assume that the magnetic field we employ, $\vec{B}$, is large
enough so that the electronic spins in the dots are aligned along the
direction of $\vec{B}$, and henceforth play no role in the subsequent
evolution. However, errors will be introduced in general if there
is a small deformation in the geometry of the loop due to
fluctuations in the control parameters, giving eventually different
phases. We can remedy this by discretising
the trajectory of the electron by allowing tunnelling between
three different dots, as shown in Fig. \ref{fig:loop}. This
structure, which eliminates statistical deformations in the
electron trajectory, is then repeated as shown in
Fig. \ref{fig:array}, and is the basis of our proposal.

By using external control we allow the activation and deactivation of
tunneling between neighbouring sites. We assume that all tunnel
couplings $t_{ij}$ between adjacent sites $i$ and $j$,
$\{i,j\}=\{1,2,3\}$, can take the same value, $J$.
In addition, we require a double dot structure in order to
implement the conditional phase gate; the purpose of this
structure is to suppress tunnelling between $1\rightarrow 3$ and
$2 \rightarrow 3$ when dot 4 is occupied. This is achieved through
the Coulomb blockade, where the Coulomb repulsion between two
electrons in each dot of the double dot is strong enough to
prevent any tunnelling towards dot 3.
In this regime, by turning on the couplings $t_{12}$,
$t_{23}$, $t_{12}$ and $t_{31}$ in succession for a sufficient time to allow
complete transition from one dot to another, it is impossible to
circulate an electron initially in dot $1$ around dots $1$, $2$,
$3$ and back to $1$. If, however, dot 4 is empty, an electron can
go around the closed path acquiring a phase due to the
Aharonov-Bohm effect. This phase is given by $\phi= A B e/\hbar$
where $A$ is the area of the triangle spanned by the dots $1$, $2$
and $3$. In both cases, at the end of this evolution, the system
returns back to the computational space where dots $2$ and $3$ are
unoccupied. We can describe this evolution, in the computational
basis, by
\begin{equation}
U= \left( \begin{array}{cccc} 1 & 0 & 0 & 0
\\
0 & e^{i \phi} & 0 & 0
\\
0 & 0 & 1 & 0
\\
0 & 0 & 0 & 1
       \end{array} \right) .
\label{cp}
\end{equation}
The resulting two-qubit gate is therefore a control-phase gate.

\section{The physical system}

We now discuss the physical conditions required in order to
implement the control procedures outlined in the previous section.
The main element that provides the controlled evolution is the
double dot. If dot 4 is occupied we would like to prohibit
tunneling towards dot 3 from dots 1 or 2. This, in effect, can be
achieved by Coulomb blockade due to strong Coulomb interaction
between electrons in dots 3 and 4. At the same time we would like
to exclude any unwanted tunneling transitions between the
different sites of the double dot in order to protect the
computational space. How these conditions are met in a realistic
setup is presented in the following.

\subsection{Coulomb blockade}

Assume that dot 4 is occupied. We would like to show that for
sufficiently strong Coulomb repulsion due to the presence of an
electron in dot 4 the Coulomb blockade effect takes place. Hence,
tunneling from, e.g. dot 1, to dot 3 will be suppressed. The
Hamiltonian for the relevant quantum dot system is given by
\begin{equation}
H= -J (a^{\dag}_1 a_3 +a_1 a_3^{\dag}) + U  a^{\dag}_3 a_3
\end{equation}
where $a, a^\dag$ are annihilation and creation operators, $U$ is
the Coulomb repulsion between the two adjacent sites 3 and 4, and
$J$ is the tunneling of an electron between two empty dots. In this
description we do not allow the
possibility of double occupation, as we assume that the barrier
between sites 3 and 4 is very large so that no tunneling can
occur. In matrix representation the Hamiltonian is given by
\begin{equation}
H=\left(\begin{array}{cc}
0 & -J \\
-J & U \\
\end{array}\right)
\end{equation}
in the basis $\ket{01} \equiv a^{\dag}_1 \ket{0}$, $\ket{10}\equiv
a^{\dag}_3 \ket{0}$ where $\ket{0}$ represents the vacuum state.
In the regime of $U>>J$, the Coulomb interaction is much larger
than the tunnelling, and we can calculate an `effective
tunnelling' parameter, $I$, felt by an electron in dot $1$. This
is related to the tunnelling rate of an electron going from  dot 1
to dot 3 when 4 is occupied. Diagonalisation of the Hamiltonian
and expansion with respect to large $U$ gives to first order
$I= J^2/U$. Hence, for large enough $U$ the tunneling from dot $1$
to dot $3$ is eliminated, as $I<<J$.

\subsection{The double dot}

As a next step we are interested in studying the regime of parameters
that allows the double dot to behave as a control device. In
particular, we demand that in the case that one side of the
quantum dot is occupied, there is no significant effective
tunneling coupling, $I$, due to the Coulomb blockade. This is
achieved by having a large Coulomb repulsion, $U$, between two
electrons occupying each site of the double dot. In order to
achieve this, the positions of the trapping minima for these
electrons are required to be close enough to each other. At the
same time these minima cannot be too close as we demand that no
significant tunnelling occurs between the two sides of the double
dot. This can also be prohibited  by maintaining a large potential
barrier between them. The regime where these conditions are
fulfilled is derived analytically in the following.

The Hamiltonian governing the double dot system is given
by~\cite{div}
\begin{eqnarray}
H&=&\int d {\mathbf{r}} \Psi^\dag ({\mathbf{r}})
\left(-\frac{\nabla^2}{2m} +V({\mathbf{r}}) \right)
\Psi({\mathbf{r}})+
\half \int\int d {\mathbf{r}} \; d{\mathbf{r'}} \; \Psi^\dag
({\mathbf{r}}) \Psi({\mathbf{r}})V_C({\mathbf{r}}-{\mathbf{r'}})
\Psi^\dag ({\mathbf{r'}}) \Psi({\mathbf{r'}}) \label{eq:Hamilt}
\end{eqnarray}
where $V({\mathbf{r}}_i)$ is the external trapping potential, which
is taken here to be quartic in the $x$ direction with its minima
separated by distance $d$, and harmonic in the
$y$ and $z$ directions
\begin{equation}
V({\mathbf{r}}_i)= \frac{m}{2} \left[ \frac{\omega_x^2}{d^2}
(x^2-\frac{d^2}{4})^2+ \omega_y^2 y^2 + \omega_z^2 z^2 \right]
\end{equation}
and $V_C$ is the Coulomb potential. For electrons well localised
in the single particle wells (harmonic wells) it is convenient to
expand the theory in terms of localised Wannier functions. The
transformed second quantised operators in the Wannier basis are
given by $\Psi({\mathbf{r}}_i)=\sum_i a_i
w({\mathbf{r}}-{\mathbf{r}}_i)$ and hence Eq. (\ref{eq:Hamilt}) becomes
\begin{equation}
H = -\sum_{ij} t_{ij} a^\dag_{i\sigma} a_{j\sigma}
+ \sum_{i i'j j'} U_{i i' j j'} a^\dag_{i \sigma}
a_{j \sigma} a^\dag_{i'\sigma'} a_{j'\sigma'}
\end{equation}
where
\begin{eqnarray}
t_{ij} &=& \int d\mathbf{r}
w^*({\mathbf{r}}-{\mathbf{r}}_i) \left(-\frac{\nabla^2}{2m}
+V({\mathbf{r}}) \right)
w({\mathbf{r}}-{\mathbf{r}}_j) \label{eq:tunnel} \\
U_{i i' j j'}  &=&\half \int d {\mathbf{r}} \int d{\mathbf{r'}} \;
V({\mathbf{r}}-{\mathbf{r'}})
w^*({\mathbf{r}}-{\mathbf{r}}_i)w({\mathbf{r'}}-{\mathbf{r}}_i')
 w({\mathbf{r}}-{\mathbf{r}}_j)
 w^*({\mathbf{r'}}-{\mathbf{r}}_j')\nonumber \\
\end{eqnarray}
For tight enough traps one can assume that the separation between
the ground state and the first excited state is large so that the
particles are well confined to the ground state in which the
potential is parabolic to good approximation. Hence the wave
functions have the form $w({\mathbf{r}})=w(x)w(y)w(z)$, each $w$
being the ground state of the harmonic potential. It is therefore
possible to calculate the tunneling and collisional couplings. In
particular for the tunneling coupling we have
\begin{equation}\label{eq:tunnel_coupling}
t_{ij}= \frac{\hbar}{2}( \omega_x + \omega_y + \omega_z ) e^{-
\alpha^2_x d^2/4 }
\end{equation}
where $\alpha^2=m\omega/\hbar$. For well localised electrons in
the quantum wells we can consider the interaction component of the
Hamiltonian taking the form of the Coulomb repulsion, $U_{ij} n_i
n_j$, where $d$ is the separation between the dots, and
$U_{ij}$ is given by
\begin{equation}\label{eq:coulomb}
U_{ij} = \frac{1}{2}\frac{{q_e}^2}{4 \pi \epsilon_0  \kappa d}
\int \int d {\mathbf{r}}  d{\mathbf{r'}} \;
|w({\mathbf{r}}-{\mathbf{r}}_j)|^2
|w({\mathbf{r}}-{\mathbf{r}}_i)|^2 = \frac{{q_e}^2}{8 \pi
\epsilon_0 \kappa d } \; e^{- \alpha ^2
 d^2/2}
\end{equation}
The conditions on the nearest-neighbour Coulomb repulsion are
stringent enough not to allow electrons to occupy adjacent dots.
We can thus safely ignore the probability of double occupation
within the same dot.

\section{Quantitative estimates and parameter regimes}

In the previous sections we established the requirements for the
double dot to provide  the controlled circulation of the electron
around the dots 1, 2 and 3. This, as we have seen, produces the
control phase gate because of the Aharonov-Bohm effect, conditional
on the presence of an electron in dot 4. In this section we want
to establish the feasibility of the above procedure by estimating
its physical requirements on the system. In particular, we want to
ensure that the time taken for an electron to make a complete
loop, i.e. the total time for the control gate, is shorter than the
decoherence time.

\begin{figure}[!htp]
\begin{center}
\includegraphics[width=18cm]{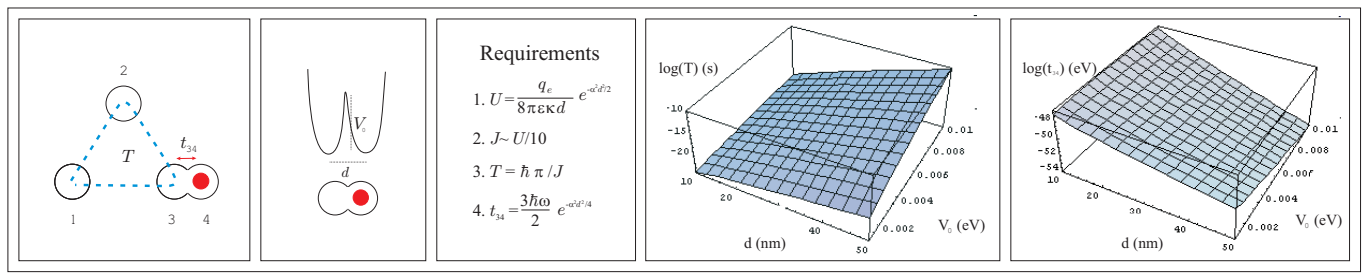}
\end{center}
\caption[chain]{Plots showing the variation of the total gate
time $T$ and of the tunnelling strength $t_{34}$ within the double
dot as a function of the barrier height $V_0$ and the distance
$d$ between the two dots. The physical setup we consider here is
quantum dots in GaAs, where the effective electron mass is $m = 0.067
m_e$ and the dielectric constant is $\kappa=13.1$. } \label{fig:plots}
\end{figure}

The time taken for an electron to tunnel to an adjacent site is
given by $t= \hbar \pi/J$. Recent experiments \cite{fuji} place a
lower bound on the decoherence time of charge qubits in
GaAs/AlGaAs heterostructure devices of the order of $ns$, which
gives us an indication of the maximum total gate time, $T=4 t $,
that can be tolerated. Since $T$ is a function of the interdot
tunnelling, it is also a function of the confinement potential,
characterised by $V_0$, the height of the barrier, and $d$, the
distance between the two dots. The distance between the dots
also determines the strength of the Coulomb repulsion, $U$,
according to Eq. (\ref{eq:coulomb}), which we needed to be much
greater than $J$
for the Coulomb blockade regime to hold. This places an upper bound on
$J$, and, in turn, a lower bound on $T$.

A plot of $T$ vs $d$ and $V_0$ is shown in Fig. \ref{fig:plots}. We
also show a plot of the intradot tunnelling in the double dot,
$t_{34}$, as a function of $d$ and $V_0$.
We can see from the plots that for the selected range of $d$ and
$V_0$, achievable with present technology \cite{kouwen}, it is
possible to obtain short gate times well below $ns$. In reality,
however, the gate time will also be limited by the finite pulse
length which controls the switching of the tunnel couplings. The
plots also show that the intradot tunnelling $t_{34}$ can also be
made small enough to be safely neglected.

\section{Discussion and Conclusions}

It is possible to manipulate the tunnel couplings in a unified
fashion to relax the stringent experimental requirement of their
individual control. We can consider for
example the case where $t_{12}$ and $t_{13}$ have a common control
procedure and $t_{23}$ is controlled independently. Indeed in this
case we can set the time for the different processes in order to
achieve the conditional phase presented in the previous section. In
particular, the interaction time has to satisfy both $Jt/\hbar=
m\pi$ and $\sqrt{2}Jt/\hbar= (2n+1)\pi$, for integers $m$ and
$n$ in order to ensure full population transfer between the
different dots in the case of dot 4 being occupied or empty. This
can be done by taking the time intervals for the duration of the
tunnel couplings in the case of the double dot being occupied or
empty to be equal, up to an integer multiple, given in the lowest
case by $m=5$ and $n=3$ respectively. This choice of integers
gives an error in the phase-gate due to timing mismatch of the
order of $1\%$.

However, our analysis does not take into account other possible
sources of errors, such as decoherence due to the coupling of the
electron to the environment. The environment can, for example, act
as a projective measurement determining the position of the
electron and thereby destroying any superposition of the electron
occupancy states of different dots. Alternatively, the environment
can destroy the phase coherence between different elements of the
superposition. For the particular analysis of the behaviour of
geometric phases under classical and quantum noise see, for
example, \cite{Vourdas}. To successfully compensate for this kind
of error, in parallel to the methods presented here, we will,
most likely, have to resort to other existing methods such as
quantum error correcting codes \cite{Steane} or error avoiding
methods such as decoherence-free subspaces \cite{Almut}.
Furthermore, another possible realisation of this scheme which may
yield a better accuracy in the manipulation of the electron around
the loop is to use a solenoid inside the loop rather than a
magnetic field~\cite{Jiannis}. Mathematically, however, this is
completely equivalent to using a uniform magnetic field, and may
be harder in practice to realise. Note finally that we can
interpret our structure as generating a single electron
circulating current (vortex) by the electron in dot $r_1$
conditional on the presence of an electron in $l_2$. This is a
very interesting physical system in its own right that could be
potentially used for other applications of quantum circuits, for
example in measuring the flux of the magnetic field by the
resulting relative phase between the two distinct possible
evolutions. A similar model with three potential wells has also
been used to describe generation of vortices in trapped
Bose-condensates \cite{Milburn}. An elaborated analysis of our
proposal including various decoherence mechanisms is therefore
very much worthwhile and will be presented elsewhere.

\acknowledgments

This work was supported by a Royal Society URF and the Schiff
Foundation.

\end{document}